\documentstyle[11pt,epsfig]{article}
\newcommand{\beq}{\begin{equation}}
\newcommand{\eeq}{\end{equation}}
\newcommand{\ba}{\begin{array}}
\newcommand{\ea}{\end{array}}
\newcommand{\EQN}{\label}
\newcommand{\dsp}{\displaystyle}
\newcommand{\api}{\frac{\alpha_s}{\pi}}
\def\bbuildrel#1_#2^#3%
  {\mathrel{\mathop{\kern 0pt#1}\limits_{#2}^{#3}}}

\begin{document}
\begin{titlepage}
\noindent
%
%
\hfill TTP93--24\\
\mbox{}
\hfill  July 1993   \\   
%
%
\vspace{0.5cm}
\begin{center}
  \begin{Large}
  \begin{bf}
Flavour Singlet ${\cal O}(\alpha_s^2 m_b^2/m_t^2)$
 Corrections to the
Partial Decay Rate $\Gamma(Z\rightarrow b\bar{b})$
   \\
  \end{bf}
  \end{Large}
%
%
  \vspace{0.8cm}
  \begin{large}
K.G.Chetyrkin    \\[3mm]
    Institute for Nuclear Research \\
    Russian Academy of Sciences\\
    60th October Anniversary Prospect 7a\\[2mm]
    Moscow 117312, Russia\\[8mm]
    A.Kwiatkowski    \\[3mm]
    Institut f\"ur Theoretische Teilchenphysik\\
    Universit\"at Karlsruhe\\[2mm]
    76128 Karlsruhe, Germany\\
  \end{large}
%
%
  \vspace{3cm}
  {\bf Abstract}
\end{center}
\begin{quotation}
\noindent
We analytically compute the flavour singlet
 ${\cal O}(\alpha_s^2 m_b^2/m_t^2)$
radiative corrections to the partial decay rate
 $\Gamma(Z\rightarrow b\bar{b})$.
These corrections arise from  anomalous
``double triangle'' diagrams
containing a  single $\gamma_5$
matrix inside each of two closed fermion loops.
They represent the next-to-leading
 term of the asymptotic expansion in the
inverse large top mass.
 As a byproduct of the calculcation
we confirm the results of B.Kniehl and J.H.K\"uhn
for the flavour singlet massless corrections
 \cite{KniKue90}
 as well as our previous
${\cal O}(\alpha_s^2 m_b^2/M_Z^2)$ result
\cite{CheKwi93}.

\end{quotation}
\end{titlepage}

\section{Introduction}
\renewcommand{\arraystretch}{2}                             %

The decay of the $Z$-boson into a bottom
 quark pair has become of
increasing interest in the past. On the
experimental side the measurement
of the partial decay rate $\Gamma(Z\rightarrow
 b\bar{b})$
represents one of the high precision
 tests of the Standard Model, which
are performed at LEP. The experimental
 uncertainty of this quantity
 $\Gamma(Z\rightarrow b\bar{b}) =
(383\pm 5.6)$ MeV is likely
 to be reduced even below 1\% in the future
\cite{Sch93}.

In order to  match this accuracy
 also on the theoretical side, a
precise knowledge of radiative corrections
 to the partial $Z$-decay rate becomes mandatory.
For this reason the influence of the
 finite bottom mass became a relevant
 topic in QCD calculations. Bottom mass
effects have conveniently been
studied by expanding radiative corrections
in $m_b^2/M_Z^2$, which is
certainly a small parameter.
 Several contributions
of the quadratic mass corrections are known
 in the literature.
Flavour non-singlet corrections were obtained
 to third order
${\cal O}(\alpha_s^3 m_b^2/M_Z^2)$ for the vector
current induced rate
\cite{CheKue90} and to second order ${\cal O}
(\alpha_s^2 m_b^2/M_Z^2)$
 to the axial vector induced rate
 \cite{CheKueKwi92}.
Flavour singlet contributions to the $Z$-width
were first discussed
for the massless case in \cite{KniKue90}. It was
 pointed out that
they arise at second order only for the axial
 induced rate, whereas
the vector contribution vanishes at this order
due to Furry's theorem.
The corresponding mass corrections were obtained
in \cite{CheKwi93}.

In this paper we focus again on the flavour singlet
 corrections to the partial $Z$-width. The
accurate  measurement of the various free
parameters of the Standard Model made it possible
to derive constraints on the
top mass. With $m_t=(158^{+30}_{-34}\pm 20 \pm 12)$ GeV
\cite{Mni93}
the top quark is expected to be roughly
twice as heavy as the $Z$-boson. Therefore one
can expect mass corrections
of order ${\cal O}(m_b^2/m_t^2)$ to be of similar
size as those of order
${\cal O}(m_b^2/M_Z^2)$.
 The inclusion
of the ${\cal O}(m_b^2/m_t^2)$ contribution
 completes the
 discussion about flavour singlet bottom mass
corrections.

Our approach for the actual calculation
of the decay rate
\beq \EQN{i1}
\Gamma^A(Z\rightarrow b\bar{b})
=\frac{G_FM_Z^3}{24\sqrt{2}\pi}N_Ca_b^2R^A
\eeq
is to employ its relation to the
absorptive part of
the corresponding axial current
correlation function
\beq \EQN{i1a}
R^A(s) = -\frac{4\pi}{s}
\mbox{\rm Im} \Pi^A_1(s+i\epsilon)
.\eeq
Here $\Pi^A_1$ arises from the Lorentz
decomposition of the 2-point
current correlator
\beq \EQN{i2}
\ba{ll}\dsp
\Pi^A_{\mu\nu}(q)
& \dsp
=i\int dx e^{iqx}\langle 0|\;T\;\Delta_{\mu}(x)
\Delta_{\nu}(0)\;|0\rangle
\\ & \dsp
= g_{\mu\nu}\Pi^A_1(q^2) + q_{\mu}q_{\nu}\Pi^A_2(q^2)
\ea\eeq
with $\Delta_{\mu} = A^t_{\mu}-A^b_{\mu}$,
 where we denote the axial vector current for the
quark of flavour $f$ with $A^f_{\mu} =
 \bar{\Psi}_f\gamma_{\mu}\gamma_5\Psi_f$.

In order to extract the $m_b^2/m_t^2$-contribution
 of the correction, we
integrate out
the heavy top quark from the
theory. This results in a power expansion
with respect to the inverse
heavy mass $1/m_t$ and leads us to the
$m_b^2/m_t^2$-term we are
interested in.

As a byproduct of the calculation we  recover the
leading and next-to-leading top dependence
of the massless flavour singlet
corrections of \cite{KniKue90} as well
 as the leading contribution
of our corresponding  massive calculation
\cite{CheKwi93}.
Whereas in \cite{CheKwi93} we made use of
the symbolic manipulation system REDUCE
\cite{Hea91}, the complexity of the calculation
now exceeds the capability
 of this software. In this work our computation
 is performed
with the help of the program MINCER \cite{LarTkaVer91}
which is designed for
massless
multiloop calculations and is based on the
symbolic manipulation package FORM
\cite{Ver91}.

This paper is organized in the following way.
In Section 2 we give a description of the
 calculation of the
${\cal O}(\alpha_s^2 m_b^2/m_t^2)$-correction.
We briefly address our  treatment of $\gamma_5$
 in $D\neq 4$ dimensions.
Further emphasis is laid on the details of the
 hard mass procedure.
Afterwards
the verification of the existing lower
order results is presented.
 The numerical discussion and conclusions
are the contents of
Section 3.

\section{Description of the Calculation}

The two-point correlation function $\Pi^A_{\mu\nu}$
is represented
 graphically by propagator-type ``light-by-light''
 Feynman diagrams,
which are characterized by two closed fermion loops
 and a purely gluonic
intermediate state. We need not consider
the diagonal pieces of
 $\Pi^A_{\mu\nu}$: The diagram with two top quark
 loops may be disregarded,
because the $Z$-boson cannot decay into top quarks
 due to the heaviness of
the top.
Furthermore, being interested in ${\cal O}
(\alpha_s^2 m_b^2/ m_t^2)$
corrections to $\Gamma(Z\rightarrow
b\bar{b})$, also the
``light'' graph (see Figure 1a) with two bottom
loops does not contribute to this order.
 We therefore have to calculate the
``heavy'' diagrams
of Figure 1b. They carry a relative factor $2$
 as compared to the
light diagram due to the top-bottom symmetry.
 The double arrows at the fermion
lines indicate the
two different diagrams with opposite
loop orientation.
\begin {figure}[t]
\begin {tabular}{ccrc}
\parbox{3cm}{
\epsfig{file=gl.eps,width=5.cm,height=5.cm}
}
&\hphantom{XXX}  & $2\times$ &
\parbox{3cm}{
\epsfig{file=ghar.eps,width=5.cm,height=5.cm}
            }
\end {tabular}
\caption { Flavour singlet diagrams contributing to
 $\Gamma(Z\rightarrow b\bar{b})$
 (thin lines: bottom, thick lines: top).}
\end{figure}
The two-point function $\Pi^A_{\mu\nu}$ of
eq.(\ref{i2}) contains the
difference between the axial currents of the
top and the bottom quark
and is therefore an anomaly free quantity.
Its absorptive part
 is also ultraviolet (UV)
finite. Although the contributions
 of the separate diagrams
are UV divergent, their singularities are
independent of the fermion mass
and compensate each other.
 At intermediate stages of the calculation
 the introduction of a regularization
procedure  therefore becomes necessary
since we calculate
the  light and the heavy diagrams separately.

 Concerning the  question of how to
define the $\gamma_5$-matrix in $D\neq
4$ dimensions, we essentially
use the definition of 't Hooft and Veltman
\cite{HooVel72}
formalized by Breitenlohner and Maison \cite{BreMai77}.
 In $D= 4-2\epsilon$ dimensions we will work
 with the generalized current
\beq\EQN{d1}
A_i^{[\mu\nu\lambda]} = \bar{\Psi}_i
\gamma^{[\mu\nu\lambda]} \Psi_i
,\eeq
where
$\gamma^{[\mu\nu\lambda]}
 = ( \gamma^{\mu}\gamma^{\nu}\gamma^{\lambda}
- \gamma^{\lambda}\gamma^{\nu}\gamma^{\mu})/2
$
and
$i=b,t$.
Taking the limit $D\rightarrow 4$ at the end
 of the calculation, when all
 divergences have been removed, we recover the
 result for the axial current
$A^{\rho}$ through
\beq\EQN{d2}
 A_{\rho} = \frac{i}{3!}\epsilon_{\rho\mu\nu\lambda}
             A^{[\mu\nu\lambda]}
.\eeq
The corresponding generalized current correlator
may be decomposed in the following way
\cite{NacWet79}
\beq \EQN{d3}
\ba{ll} \Pi^{[\mu\nu\lambda]}_{[\mu'\nu'\lambda']}
& = \dsp i \int dx e^{iqx}
\langle 0|\;
T\; A^{[\mu\nu\lambda]}(x)
A^{[\mu'\nu'\lambda']}(0)\; |0 \rangle
\\ &
 = \dsp {[\mu\nu\lambda] \atop [\mu'\nu'\lambda']}
{\rm P}_1(q^2)
+g^{[\mu}_{[\mu'}g^{\nu}_{\nu'}
q^{\lambda ]}q_{\lambda']}
   {\rm P}_2(q^2)  ,  \ea
\eeq
where
$
 {[\mu\nu\lambda] \atop [\mu'\nu'\lambda']}
= \frac{1}{3!} {\rm det}(g_{\alpha\alpha'})$ with
$
\alpha=\mu,\nu,\lambda
$
and
$
 \alpha'=\mu',\nu',\lambda'
$
and where $g^{[\mu}_{[\mu'}g^{\nu}_{\nu'}
q^{\lambda ]}q_{\lambda']}$
is  fully antisymmetrized with respect
to $[\mu\nu\lambda]$
and $[\mu'\nu'\lambda']$ separately.
It is  convenient to work with the
contracted tensors
$
P_L=\Pi^{[\mu\nu\lambda]}_{[\mu\nu\lambda]}
$
and
$
P_Q= q^{\lambda}q^{\lambda'}
\Pi^{[\mu\nu\lambda]}_{[\mu\nu\lambda']}
$.
It turns out that the spectral function
$\Pi^A_1$, which enters
the $Z$-decay rate, is already completely
 determined by $P_Q$:
\beq\EQN{d4}
\Pi^A_1 =  \frac{1}{6}
\frac{P_Q}{q^2}
.\eeq

The calculation of the heavy diagrams of
Figure 1 proceeds as follows.
We apply a hard mass expansion in the
 inverse heavy top mass $1/m_t$.
In practice this means that the Feynman
 integral $\langle\Gamma \rangle
(q^2)$ corresponding to a Feynman diagram
$\Gamma$ is represented by an
asymptotic expansion
 \cite{GorLar87,PivTka84,Smi90,Smi91}):
\beq
\EQN{d5}
<\Gamma>(q^2)
\bbuildrel{=\!=\!=}_{{\scriptstyle{m_t\to\infty}}}^{}
\sum_\gamma  C^{(t)}_\gamma \star
\langle\Gamma/\gamma\rangle^{\mbox{\rm
 \scriptsize eff}}
.\eeq
Thereby the hard subgraphs $\gamma$
 of $\langle \Gamma\rangle$ are reduced to
effective vertices such that  the diagram
$\langle\Gamma/\gamma\rangle^{\mbox{\rm
\scriptsize eff}}$
is built up only from the (light)
particles of the effective theory.
The subintegrals $C^{(t)}_\gamma$,
which contain the whole top dependence, are
expanded in a formal (multidimensional)
 Taylor series with respect
to their external momenta and the light
bottom mass. This Taylor
expansion is then inserted into the effective
vertex and the whole
expression is evaluated.

In our case we can identify two possible
hard subgraphs in our diagrams
as is shown in Figure 2.
\begin {figure}[t]
\begin {tabular}{ccccc}
\parbox{3cm}{
\epsfig{file=gh.eps,width=4.cm,height=4.cm}
}
& $\longrightarrow$ &
\parbox{3cm}{
\epsfig{file=ct1.eps,width=4.cm,height=4.cm}
            }
& $\star$ &
\parbox{3cm}{
\epsfig{file=gh1.eps,width=4.cm,height=4.cm}
            }
\\
& $+$ &
\parbox{3cm}{
\epsfig{file=ct2.eps,width=4.cm,height=4.cm}
            }
& $\star$ &
\parbox{3cm}{
\epsfig{file=gh2.eps,width=4.cm,height=4.cm}
       }
\end {tabular}
\caption {The hard mass procedure.}
\end{figure}
The hard mass procedure results in a
 power series in the inverse top mass.
The leading order of this expansion
 has the top mass dimension $m_t^0$
and receives its contribution only
 from the two-loop subgraph $C^{(t)}_1$.
For the next-to-leading order $\sim
 1/m_t^2$ the second order terms in the
Taylor expansion of $C^{(t)}_1$ as
well as the  contributions of the
 one-loop subgraph $C^{(t)}_2$ are
needed. Since the Taylor series
is evaluated at vanishing external momenta,
the subgraphs are computed by
 solving two- and one-loop massive
tadpole integrals respectively.

After having done this, the problem is
reduced to the calculation of
one- and two-loop diagrams, which still
contain the bottom quark mass.
We perform a subsequent expansion
 in $m_b^2$.
Picking out the quadratic mass terms
in order to obtain the
order ${\cal O}(\alpha_s^2 m_b^2/ m_t^2)$
corrections to the $Z$-decay rate,
we are left with massless  propagator
 type diagrams, which we compute
 with the help
of the multiloop program MINCER.
For the order ${\cal O}
(\alpha_s^2 m_b^2/m_t^2)$ corrections
we arrive at the following result:
\beq\EQN{d6}
R^A = - 10 \left( \frac{\alpha_s}{\pi}\right)^2
\frac{\bar{m}_b^2}{m_t^2}\left[ \frac{8}{81}-
\frac{1}{54}\ln\frac{s}{m_t^2}
\right]
\eeq

\vspace{2ex}
Before closing this section let us add
some additional details and remarks.
In deriving the result of eq.(\ref{d6})
 we have isolated the next-to-leading
term in the heavy mass expansion and
 also concentrated on the
quadratic contribution of the bottom
 mass. As a byproduct of this
calculation we automatically obtain
also the lower order pieces
with respect to the $m_b^2$-expansion
 as well as the $1/m_t$-power
series. Therefore we are able to
 compare the outcome of the
calculation not only with our
 previous result for the
quadratic mass corrections for
 the flavour singlet diagrams
\cite{CheKwi93}, but also
with the result of \cite{KniKue90}
 for the massless case.
In order to verify the latter
 corrections, we have to extend our consideration
and include the light diagram of
 Figure 1. For this diagram we  perform
an expansion in $m_b^2$ and compute
 the resulting massless
three-loop integrals with the program MINCER.
We present the intermediate results
for the absorptive parts  of the
 various contributions,
namely of the light diagram $\Gamma_\ell$
and of the hard diagrams, separated
according to
the two-loop hard subgraph
($\Gamma_{h1}$) and the one-loop hard
subgraph ($\Gamma_{h2}$):
\beq\EQN{d7}
\ba{ll}\dsp
R^A(\Gamma_\ell) =
& \dsp
\left( -\frac{1}{3}\right)\left(
\frac{\alpha_s}{\pi}\right)^2
 \left[ \frac{3}{2}\frac{1}{\epsilon}
+\frac{25}{2} - \frac{9}{2}\ln\frac{s}
{\mu^2}\right]
\\ \dsp
& \dsp
 -6\frac{\bar{m}_b^2}{s}\left(
\frac{\alpha_s}{\pi}\right)^2
 \left[ -\frac{1}{2}\frac{1}{\epsilon}
-\frac{15}{4} + \frac{3}{2}\ln\frac{s}
{\mu^2}\right]
\ea\eeq
\beq\EQN{d8}
\ba{ll}\dsp
R^A(\Gamma_{h1}) =
& \dsp
 \left(-\frac{1}{3}\right)
\left( \frac{\alpha_s}{\pi}\right)^2
 \left[- \frac{3}{2}\frac{1}{\epsilon}
-\frac{13}{4} +\frac{3}{2}\ln\frac{s}{\mu^2}
- 3\ln\frac{\mu^2}{m_t^2}
 -\frac{10}{27}\frac{s}{m_t^2}
\right]
\\ \dsp
& \dsp
 -6\frac{\bar{m}_b^2}{s}\left(
 \frac{\alpha_s}{\pi}\right)^2
 \left[ \frac{1}{2}\frac{1}{\epsilon}
+\frac{3}{4} - \frac{1}{2}
\ln\frac{s}{\mu^2}
+\ln\frac{\mu^2}{m_t^2}
\right]
\\ \dsp
& \dsp
 -10\frac{\bar{m}_b^2}{m_t^2}\left(
 \frac{\alpha_s}{\pi}\right)^2
 \left[ -\frac{1}{54}\frac{1}{\epsilon}
+\frac{7}{324} + \frac{1}{54}
\ln\frac{s}{\mu^2}
- \frac{1}{27}\ln\frac{\mu^2}{m_t^2}
\right]
\ea \eeq
\beq\EQN{d9}
\ba{ll}\dsp
R^A(\Gamma_{h2}) =
& \dsp
 \left(-\frac{1}{3}\right)\left(
\frac{\alpha_s}{\pi}\right)^2
 \frac{1}{9}
\frac{s}{m_t^2}
\\ \dsp
& \dsp
 -10\frac{\bar{m}_b^2}{m_t^2}
\left( \frac{\alpha_s}{\pi}\right)^2
 \left[ \frac{1}{54}\frac{1}{\epsilon}
+\frac{25}{324} -
 \frac{1}{27}\ln\frac{s}{\mu^2}
+\frac{1}{54}\ln\frac{\mu^2}{m_t^2}
\right]
\ea \eeq
One observes that the pole terms of the
leading order cancel in the sum of
the  diagrams $\Gamma_\ell$ and
$\Gamma_{h1}$, whereas those of the
 next-to-leading contribution compensate
 each other in the combination
of $\Gamma_{h1}$ and
$\Gamma_{h2}$. In a similar way
the $\mu^2$-dependence vanishes
in the sum of the logarithms.
Finally  the result is
gauge invariant. The independence of
each of the three contributions
on the QCD gauge parameter serves as
a further check of the calculation.
We arrive at the following  result:
\beq\EQN{d10}
\ba{ll} \dsp
R^A_{singlet} =
& \dsp
\left(-\frac{1}{3}\right)\left(
\frac{\alpha_s}{\pi}\right)^2
\left[ \frac{37}{4} - 3 \ln\frac{s}{m_t^2}
- \frac{7}{27}\frac{s}{m_t^2}
\right]
\\ & \dsp
 -6\frac{\bar{m}_b^2}{s}\left(
 \frac{\alpha_s}{\pi}\right)^2
 \left[ -3+ \ln\frac{s}{m_t^2}
\right]
\\
& \dsp
 -10\frac{\bar{m}_b^2}{m_t^2}\left(
\frac{\alpha_s}{\pi}\right)^2
 \left[
\frac{8}{81} - \frac{1}{54}\ln\frac{s}{m_t^2}
\right]
.\ea\eeq
The first line of this equation
reproduces the result of
B.Kniehl and J.H.K\"uhn \cite{KniKue90}.
The second line verifies
our previous calculation in
\cite{CheKwi93}, and the third line
represents the new contribution.

\section{Discussion}
Let us first discuss
 the numerical size of our result for
the flavour singlet ${\cal O}(
\alpha_s^2 m_b^2/m_t^2)$ corrections to the
partial decay rate of the $Z$-boson.
Collecting the complete mass corrections
 for the axial vector induced
rate we obtain
\beq\EQN{r1}
\ba{ll} \dsp
R^A =
& \dsp
1 - 6 \frac{\bar{m}_b^2}{s}
\left( 1 + \frac{11}{3}\left(
\frac{\alpha_s}{\pi}\right)
+\left( \frac{\alpha_s}{\pi}
\right)^2\left[ 11.286
+ \ln\frac{s}{m_t^2}\right]
\right)
\\ &\dsp
\hphantom{1} - 10 \frac{\bar{m}_b^2}{m_t^2}
\left( \frac{\alpha_s}{\pi}\right)^2\left[
\frac{8}{81}
-\frac{1}{54} \ln\frac{s}{m_t^2}\right]
.\ea\eeq
We compare the size of the corrections
of  the orders ${\cal O}(\alpha_s^2 m_b^2/M_Z^2)$
and ${\cal O}(\alpha_s^2 m_b^2/m_t^2)$
  in Table 1
for different $m_t$. The numbers are based
 on $\Lambda_{QCD}=400$ MeV and
an on-shell mass of $m_b=4.72$ Gev
\cite{DomPav92}, which corresponds to
$\alpha_s(M_Z^2)= 0.131$ and
 $\bar{m}_b(M_Z^2)=2.61$ GeV.
One can observe that
the order ${\cal O}(\alpha_s^2 m_b^2/m_t^2)$
corrections carry opposite
sign as the ${\cal O}(\alpha_s^2 m_b^2/M_Z^2)$
 flavour singlet corrections.
In size  they are smaller about one order
of magnitude.
Due to their polynomial $1/m_t^2$ behavior
they are much more sensitive
on the top mass. However,
the complete flavour singlet mass  corrections
 amount to only  $\sim 0.03$ permille
and are too small
for LEP to be sensitive
 on the top mass
from flavour singlet QCD effects.

\begin{table}
\renewcommand{\arraystretch}{1}                             %
{\bf Table 1:} Flavour singlet mass
corrections of  $R^A.$

\vspace{2ex}

\begin{tabular}{|c||c|c||c|}\hline
$m_t$ & $R^A(\bar{m}_b^2/M_Z^2)$ & $R^A
(\bar{m}_b^2/m_t^2)$ & $\sum$
\\
in GeV & $\cdot 10^5$ & $\cdot 10^5$
& $\cdot 10^5$ \\
    \hline
110 & 2.87 & -0.103 & 2.76 \\
130 & 3.16 & -0.078 & 3.08 \\
150 & 3.40 & -0.061 & 3.33 \\
170 & 3.61 & -0.050 & 3.56 \\ \hline
\end{tabular}
\renewcommand{\arraystretch}{2}                             %
\vspace{2ex}
\end{table}

Together with
 the flavour non-singlet contribution the
second order mass corrections are just
below $-0.1$ permille in magnitude
(see column 6 of Table 2).
The size of the  quadratic mass
 corrections originating from
the separate orders in the coupling
 constant are presented in
Table 2 as well.

\begin{table}
\renewcommand{\arraystretch}{1}                             %
{\bf Table 2:} Quadratic
 mass corrections of $R^A.$

\vspace{2ex}

\begin{tabular}{|c|c|c||c|c|c||c|}\hline
$\Lambda_{QCD}$ & $\alpha_s(M_Z^2)$
& $\bar{m}_b(M_Z^2)$ &
$R^A(\alpha_s^0 \bar{m}_b^2)$ & $R^A(\alpha_s \bar{m}_b^2)$
& $R^A(\alpha_s^2 \bar{m}_b^2)$ &
$\sum$
\\
in MeV & & in GeV & $\cdot 10^3$  &
$\cdot 10^3$ & $\cdot 10^3$ & $\cdot 10^3$
 \\     \hline
200 & 0.117 & 2.94 & -6.24 & -0.85 & -0.092 & -7.34 \\
400 & 0.131 & 2.61 & -4.90 & -0.74 & -0.090 & -5.74 \\
650 & 0.143 & 2.27 & -3.70 & -0.62 & -0.081 & -4.40 \\
\hline
\end{tabular}

\renewcommand{\arraystretch}{2}                             %
\vspace{2ex}
\end{table}

For completeness we finally reproduce the
complete formula including all presently known
QCD corrections to the partial $Z$-decay rate.
\beq   \EQN{r2}
\ba{ll}
\dsp \Gamma(Z\rightarrow b\bar{b})
& \dsp =  \Gamma^V + \Gamma^A
\\ & \dsp = \frac{G_F M_Z^3}{24\sqrt{2}\pi}
         N_C (v_b^2 R^V + a_b^2 R^A),
\ea
\eeq
where
\beq  \EQN{r3}
\ba{ll}
R^V & \dsp = 1 + \api +
\left[ 1.41 - \frac{8}{135}r\ln(4r) +
\frac{176}{675}r \right]
 \left(\api\right)^2
\\ & \dsp
 - 12.76 \left(\api\right)^3
-\frac{\sum_f v_f}{3 v_b} 1.24 \left(\api\right)^3
\\ & \dsp
+ 12 \frac{\bar{m}_b^2}{s} \api \left\{ 1 + 8.7 \api
           + 45.15  \left(\api\right)^2  \right\}
\\ \dsp
R^A & \dsp = 1 + \api +
\left[ 1.41 - \frac{8}{135}r\ln(4r) +
\frac{176}{675}r
+ \frac{1}{3}{\cal I}(r)
 \right] \left(\api\right)^2
\\ & \dsp
- 6 \frac{\bar{m}_b^2}{s}  \left\{ 1 + \frac{11}{3} \api
           + \left[ 11.286 + \ln \frac{M_Z^2}{m_t^2} \right]
\left(\api\right)^2  \right\}.
\\ &\dsp
 - 10 \frac{\bar{m}_b^2}{m_t^2}
\left( \frac{\alpha_s}{\pi}\right)^2\left\{
\frac{8}{81}
-\frac{1}{54} \ln\frac{s}{m_t^2}\right\}
.\ea
\eeq
Here we have included the massless \cite{m0,m1} as well
as the quadratic mass corrections.
The term  ${\cal I}(r)$ with $r=M_Z^2/4m_t^2$
 represents the flavour singlet contribution
of the triangle diagrams for the massless case
  \cite{KniKue90} and reads
\beq \EQN{r4}
{\cal I}(x) = -9.250 + 1.037 x + 0.632 x^2 + 6 \ln (2\sqrt{x})
.\eeq
The other terms depending on the top mass through $r$
originate from non singlet double bubble diagrams
with a virtual top loop and have been calculated
in \cite{Che93}.

We finally conclude that in this work
we have determined the
 complete flavour singlet contribution
 to the quadratic mass corrections for the
axial vector induced partial decay
rate of the $Z$-boson
into bottom quarks. As a byproduct we
 could verify previous results
which were obtained earlier for the
 flavour singlet diagrams.

QCD corrections for the decay rate
$\Gamma(Z\rightarrow b\bar{b})$ are well established
by now. They are known  to a precision
which is sufficiently accurate in view of the
experimental uncertainities at LEP.

{\bf Acknowledgment}

We would like to thank J.H.K\"uhn for helpful discussions.

\end{document}